\documentclass[12pt]{iopart}
\usepackage{hyperref}
\usepackage{lscape}
\usepackage{multirow}
\usepackage{graphicx}
\usepackage{float}

\usepackage{xcolor}
\usepackage{verbatim}
\usepackage{caption}
\usepackage{siunitx}
\newcommand{\rmnum}[1]{\romannumeral #1 }

\begin{document}

\title[Optimizing the Optimizer: Decomposition Techniques for Quantum Annealing]{Optimizing the Optimizer: Decomposition Techniques for Quantum Annealing}

\author{Gideon Bass$^1$, Max Henderson$^2$, Joshua Heath$^1$, \\ Joseph Dulny III$^1$}
\address{$^1$ Booz Allen Hamilton, Washington, DC \\ $^2$ Rigetti Computing, Berkeley, CA}
\ead{Bass\_Gideon@bah.com}

	\begin{abstract}
		Although quantum computing hardware has evolved significantly in recent years, spurred by increasing industrial and government interest, the size limitation of current generation quantum computers remains an obstacle when applying these devices to relevant, real-world problems. In order to effectively exploit the potential benefits of quantum computing, heterogeneous approaches that combine both classical and quantum computing techniques are needed. In this work, we explore multiple heterogeneous approaches to solving multiple industry-relevant benchmark problems in order to understand how best to leverage quantum computers given current constraints. Our results indicate: that solver performance is highly dependent on the structure (size and edge density) of the problem graph; that reusing a single fixed problem embedding, as opposed to dynamically searching for problem embeddings, is key to avoiding computational bottlenecks; that solutions of better quality are produced by algorithms that iteratively propagate the influence that solving an individual sub-problem has to the remainder of the larger problem; and that the Qbsolv algorithm (which implements the aforementioned techniques) is, at this time, the state-of-the-art in producing quality solutions, in a timely fashion, to a variety of theoretical and real-world problems too large to directly embed onto a quantum annealing device.

	\end{abstract}


	\section{Introduction}
	\label{introduction}

		The field of quantum computing is entering the so-called "NISQ-era" of Noisy, Intermediate-Scale Quantum computers \cite{preskill_nisq}. While the NISQ-era is expected to yield quantum computers that can surpass certain computational benchmarks, such as quantum supremacy \cite{Preskill2012}, their applicability to real-world problems will still be constrained by current generation hardware limitations. Currently, the largest (in terms of qubit count) commercially available quantum computing hardware device is the D-Wave 2000Q quantum annealer \cite{Systems2018}, which has 2048 qubits when functioning at full capacity and is designed to solve a particular class of difficult optimization problems \cite{Farhi_2001, Garey_1990} using a quantum annealing algorithm \cite{Brooke_1999, Farhi_2001, Finnila_1994, Kadowaki_1998, Santoro_2002, Santoro_2006}. While this device showcases the rapid hardware advances made in the field from the first 128-qubit commercially available device offered in 2011, current hardware is vastly exceeded by the thousands or millions of variables that real-world industrial problems often require. 
		
		A common workaround to this limitation has been to investigate the scalability of quantum applications using small exemplar problems. Performance results generated from these investigations are then used to extrapolate algorithm performance on larger devices. While scalability studies are vital for understanding longer-term quantum computing viability, they do not answer the question: ``What is the optimal way to use currently available quantum devices to solve real problems today?"

		A promising path forward for mitigating quantum hardware restrictions is to explore heterogeneous classical-quantum computing solutions. Heterogeneous approaches harness the strengths of both classical and quantum systems for solving difficult problems by utilizing the quantum component as a specialized sub-processor within a larger classical algorithm. Several previous projects have evaluated heterogeneous approaches using existing quantum annealing devices \cite{Bass2018, bian2014, Bian2016, Chapuis2019, Djidjev2016}. However, many of these algorithms were developed for specific problems - a paradigm that is exemplified in \cite{Ajagekar2019}. That work, while a thorough examination of problem decomposition techniques across multiple problems, focused on developing targeted decomposition algorithms, each of which was aimed at and evaluated on one of the particular problems examined. An exhaustive analysis of the performance of various heterogeneous techniques - both existing and new - when solving a variety of diverse problems is currently absent from the literature.

		The goal of this work is to provide a more comprehensive understanding of best practices for heterogeneous quantum annealing algorithms to solve industrially-relevant optimization problems with different problem graph characteristics. We accomplish this through a multivariate evaluation of four heterogeneous methods on four problems, each of which is relevant to a real-world application and has a unique underlying graph structure. Our work thus stands out as unique from previous in developing and comparing generalized heterogeneous algorithms capable of solving any arbitrary quadratic unconstrained binary optimization (QUBO) problem - the native problem format of current quantum annealing devices.
		
		With the first true ``quantum benefit" occurring for a highly specialized application \cite{Arute2019}, determining how to optimize for every level of the quantum computing stack is pivotal. While heterogeneous solutions will likely not provide exponential quantum speedups in the near term, they have the potential to offer sub-exponential improvements in algorithm performance for valuable real-world problems. This is not negligible, as even minor algorithmic improvements as a result of leveraging quantum could translate to notable industrial or military benefit, subsequently spurring resource devotion to quantum computing research. We believe that, for quantum computing, there will always be a virtuous cycle between scientific progress and industrial/defense applications, and lines of research investigating these applications are powerful for moving the field forward.
		
		This work is organized as follows: In section \ref{sec:framework}, we discuss the overall experimental set-up, including the specifics of the problems and heterogeneous techniques used. In section \ref{sec:results} we present and discuss the results of the many experimental runs we performed. Finally, section \ref{sec:conclusion} summarizes our conclusions and provides overall commentary on the current state-of-the-art in heterogeneous optimization, as well as ideas for improvements to current and future techniques. 

	\section{Quantum Heterogeneous Optimization Techniques (QHOTs) framework} \label{sec:framework}
	
		\subsection{Framework Overview and Abstract Class Descriptions} \label{abstract_class_description}
		
			To accomplish our goals of exploring and fairly comparing the performance of different heterogeneous optimization techniques when solving different types of problems, we developed an extendable, object-oriented software framework, whose design is inspired by the following problem-solving workflow:
			\begin{enumerate}
				\item a type of \textit{ProblemGenerator} is initialized;
				\item a specific \textit{QuboProblem} instance is generated using the generator; and
				\item the problem instance is submitted to and solved by a \textit{HeterogeneousSolver}. 
			\end{enumerate}
			
			These three components form the backbone of the QHOTs framework and are each implemented in a separate abstract class which establishes concrete class implementation requirements and collectively define the integration points between software components. The framework is easily extendable as new types of problems and heterogeneous techniques can be added if they satisfy integration point and class requirements.
			
			Each optimization problem solved using the QHOTs framework is first mapped to a QUBO problem representation. By definition, a QUBO problem has the following formulation:
				\begin{center}
					$E(S) = \displaystyle\sum_{i} a_i q_i + \displaystyle\sum_{i<j} b_{i,j} q_i q_j,$
				\end{center}
			where the energy function $E$ evaluates a solution vector $S$ which is a set of binary variables $S = \{q_1, q_2, ... , q_N\}$, wherein each $q_i$ represents the measured value of a logical qubit. Mapping an optimization problem into the QUBO format involves calculating the appropriate values of the $a_i$'s (known as the \textit{qubit biases}) and $b_{i,j}$'s (known as the \textit{coupling strengths}) of $E$ such that a solution $S$ that minimizes $E$ will represent an optimal solution to the original non-QUBO optimization problem formulation. 
				
			Once a ProblemGenerator has instantiated a QuboProblem, a solution is determined by submitting the problem to a HeterogeneousSolver. The HeterogeneousSolver uses a mix of classical techniques and quantum annealing to solve the overall optimization problem, with the one commonality being that the approach prepares a set of sub-QUBOs (QUBO problems that are reduced versions of the full problem) which are embedded on and solved directly by a quantum annealing device.

		\subsection{Optimization Problems} \label{concrete_qubo_problems_and_generators}
			
			In this section, we describe key aspects of each problem type evaluated in this work.

			\subsubsection{Random QUBO with time-domain structure:} 
				Many optimization problems have inherent time-based constraints. When mapping such time-constrained problems to the QUBO form, the presence of these time constraints results in the underlying problem graph having a structure similar to that of some deep neural architectures - specifically, the Deep Boltzmann Machine \cite{Salakhutdinov2009}. 
				
				Because of this similarity, problems of this type in this work are referred to as Deep Boltzmann Graph (DBG) problems. In these problems, we simulate time-domain structure by grouping variables into layers, representing different moments in time, and forming constraints (non-zero coupling strengths) between variables across some finite number of layers (time steps). With a strong time-domain constraint, variables are restricted to only having non-zero coupling strengths with variables within adjacent layers. Loosening this constraint to allow couplings between variables two or more layers apart allows the emphasis of time to be steadily lowered. At the extreme, when any two variables can be coupled, random unstructured QUBOs with up to fully-connected problem graphs can be generated.
				
				The following parameters are defined for the DBG problem, with the respective values explored in this work provided in parentheses:
				\begin{itemize}
					\item \textit{number\_of\_layers}: Number of total layers in the graph architecture, where each layer represents a moment in time (25, 50).
					\item \textit{nodes\_per\_layer}: Number of nodes per layer (20).
					\item \textit{max\_connectivity\_range\_layer}: Connections will never be formed between nodes that are more than max\_connectivity\_range\_layer layers apart (1, 2, 5, 10, 25, 50).
					\item \textit{connectivity\_probability}: Probability of forming a connection (constraint) between two nodes (including nodes within the same layer), if the number of layers between the nodes is less than or equal to max\_connectivity\_range\_layer. The edge weight, i.e. the coupling strength ($b_{i,j}$), of each constraint is set to 1 (0.1, 0.25, 0.5, 1).
					\item \textit{average\_node\_value}: Average value awarded for each node activated. The negation of the ``award value" of a node is the bias ($a_i$) of the qubit represented by that node in the final QUBO definiton (0.1).
					\item \textit{optimization\_type}: Indicates how ``award values" are determined. If set to ``constant", all nodes will have an award value of average\_node\_value. If set to ``random", each node will have a random award value sampled uniformly from the range [0, 2 $\cdot$ average\_node\_value] (``constant", ``random").
				\end{itemize}
					
			\subsubsection{Symmetric Travelling Salesman Problem:}
				One of the most well known combinatorial optimization problems is the Travelling Salesman Problem (TSP). Given an undirected graph with weighted edges, an optimal solution to the Symmetric TSP is a path through all graph vertices such that each vertex is visited exactly once and the sum of the weights of edges traversed in the path is minimal. In this work, we implement the QUBO formulation for the Symmetric TSP given in \cite{Lucas2014}.
				
				We explore the TSP on two types of graphs. The first, ``Random", is fully-connected graphs defined on 11, 20, 50, or 100 vertices, with edge weights being randomly sampled from the range [1,10]. The second type of graph is also fully connected, with each graph being derived from a unique benchmark TSP instance downloaded from TSPLIB \cite{Reinelt1995}.

			\subsubsection{Satellite sub-Constellation Assignment:} \label{sec:Satellite}
				The Satellite sub-Constellation Assignment (SCA) problem involves partitioning a set, referred to as a ``constellation", of $N$ satellites into $k$ smaller ``sub-constellations." Each SCA problem instance is defined by $N$, $k$, and the ``coverage" of each possible sub-constellation, defined as the percentage of time that an Earth region of interest is within signal range of at least one satellite within that sub-constellation. The goal is then to assign each satellite to exactly one sub-constellation such that the total sum of the coverages of each of the sub-constellations is maximized. This problem can be cast as a weighted k-clique problem, which can then be reformulated as a QUBO.
				
				In weighted k-clique problem formulation, each possible sub-constellation is represented by a single node in a graph, with a node weight equal to the coverage of the sub-constellation that it represents. An edge is present between two nodes if the two sets of satellites are disjoint, i.e. there is no satellite that is in both sub-constellations. Within the resulting graph formulation, a clique of size $k$ represents $k$ disjoint sub-constellations. The goal is to find the clique of size $k$ that maximizes the sum of node weights. For a description of the process by which the weighted k-clique problem can be formulated as a QUBO, see \cite{Bass2018}.
				
				As with all types of problems explored within this work, the size of the problem is a serious consideration. This is especially true of the real-world SCA problem, which contains over one million nodes when formulated as a weighted k-clique problem. Because the QUBO representation of the weighted k-clique problem requires a fully-connected problem graph (i.e. a non-zero coupling strength exists for every pair of qubits), the problem requires over one million logical qubits and over $10^{14}$ unique coupling strengths. Simply generating the problem coefficients required traditional memory requirements greater than the classical computing resources allocated for this work. To mitigate these memory issues, it was observed \cite{Bass2018} that solutions of comparable quality could be found by ignoring sub-constellations with coverages below a given threshold. Empirically, we found that devising a threshold that removed all but the 99th percentile of sub-constellation coverages shrunk the problem to a more manageable size. Additionally, for the 27-satellite problem instance explored in this work, we found that limiting sub-constellation size to 3, 4, or 5 and setting $k$ to 7 produced solutions of good quality. While still too large to directly embed on a quantum annealer, the reduced problem was now small enough to be easily manipulated in classical memory by our QHOTs.

			\subsubsection{Military Maintenance Planning and Scheduling: } 
				Another example of a real-world optimization problem that we explore in this work is the Maintenance Workload Problem (MWP) introduced in \cite{Squires2014}, a variant of the job shop scheduling problem. Given a set of items that require repairing, the MWP asks for the most cost-effective scheduling of these items at repair facilities that satisfies the following constraints:
				\begin{enumerate}
					\item Each item originates from a specified location during a specific week in time (release date) and must arrive, repaired, at a specified location before a specific week (due date).
					\item Each item must be shipped from its origin to the selected repair facility, and from this repair facility to its final destination, with the financial and temporal costs of these shipments being considered.
					\item During a given week, each repair facility has a limited number of worker-hours that can be used to repair items.
					\item Each repair is of a certain type and must be completed (without preemption) at a facility that can perform that type of repair, where the length (in number of weeks), cost, and required worker-hours per week for the repair are unique to the facility.
				\end{enumerate}
			
				In order to cast the MWP as a QUBO problem, we represent a potential solution (scheduling) by the combination of values taken by binary variables $x_{(R_i,L_j,T_k)}$, where $R = \{R_1, ..., R_X\}$ is the set of all repairs, $L = \{L_1, ..., L_Y\}$ is the set of all repair facilities, and $T = \{T_1, ..., T_Z\}$ is the set of all weeks between the earliest release date and the latest due date of the repairs in $R$. In this definition, $x_{(R_i,L_j,T_k)} = 1$ if repair $i$ is going to be completed at repair facility $j$ and will begin to be completed (i.e. shipped to $j$) on week $k$, and $x_{(R_i,L_j,T_k)} = 0$ otherwise. 
				
				The mathematical formulation of each of the aforementioned constraints is found in \cite{Squires2014}. To map this formulation to a QUBO, we first modified each constraint such that a solution that satisfies a given constraint would minimize the QUBO definition of that constraint. These individual constraints were then weighted and combined in an intelligent way to form the holistic QUBO formulation of the MWP. This QUBO requires on the order of $R \cdot L \cdot T$ logical qubits, though the actual number may be less because $x_{(R_i,L_j,T_k)}$ will always be 0 for combinations of repair/repair\_facility/start\_date that are impossible. 
				
				The original MWP was defined on 1 of 12 unique sets of repairs, with each set being referred to as an ``equipment group". Due to both costly qubit requirements and time limitations, we limit our exploration of the MWP to the instance defined on the smallest group (equipment group 1).
				
		\subsection{Quantum Heterogeneous Optimization Techniques (QHOTs)}
			In this section, we detail key algorithmic steps that define each heterogeneous technique investigated within this work. Each algorithm implemented is reusable in that it can be used to solve any QUBO problem, including each of the problems described in \ref{concrete_qubo_problems_and_generators}. 

			\subsubsection{Principal Component Decomposition (PCD):} 
\begin{figure}[H]
					\begin{center}
						\includegraphics[width= 0.8\linewidth]{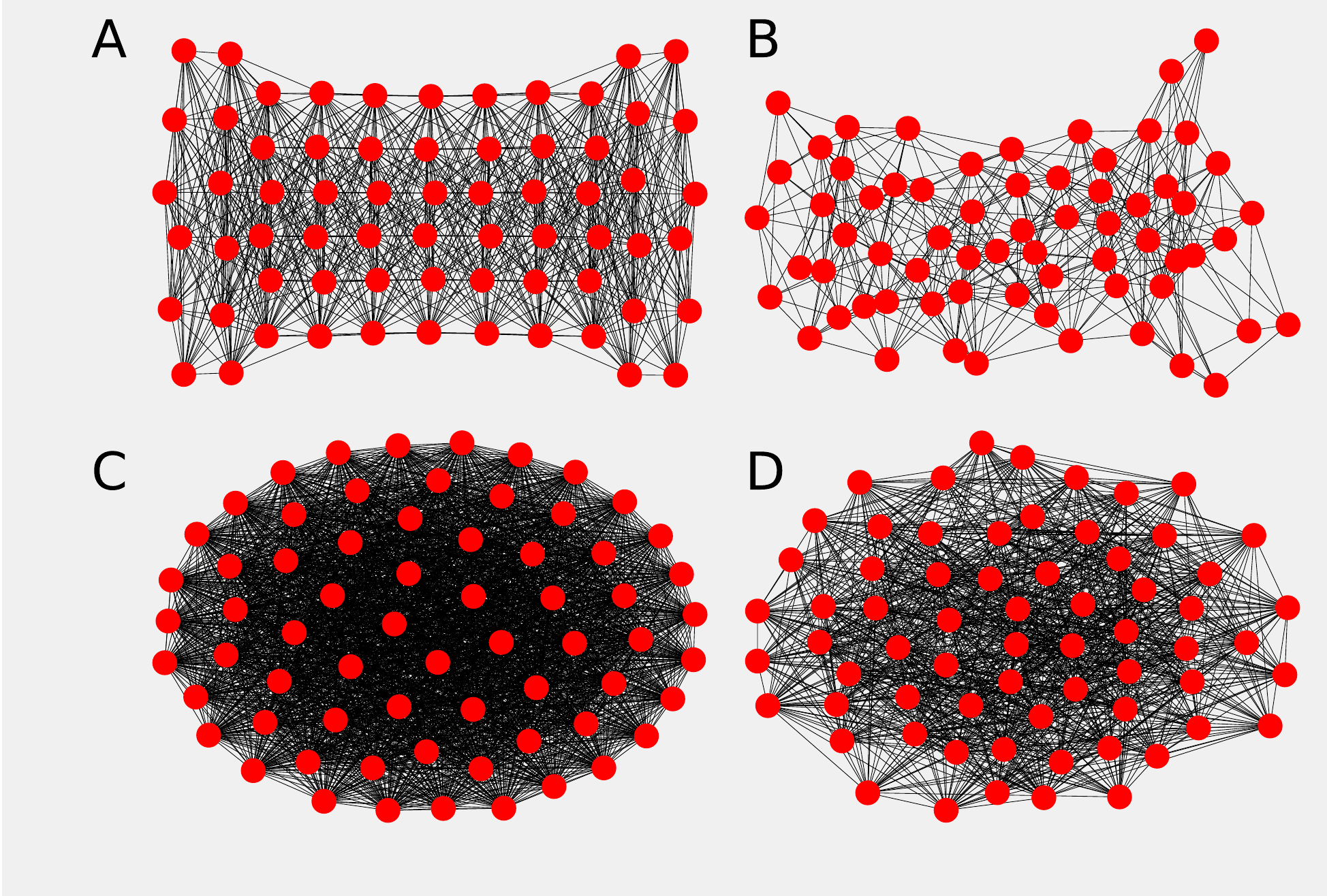}
						\caption{Final node positions determined using the PCD algorithm for 4 example DBG problem graphs defined on 11 layers of 6 nodes each. In graphs A and B, couplings can occur between nodes that are within two layers of each other; this restriction is non-existent in graphs C and D. For those couplings that can occur in these graphs, there is a 0.5 probability that a coupling between two variables does occur in B and D, and a probability of 1 for A and C. As seen in A and B, the PCD algorithm is able to regenerate the original time-domain structural patterns of the input problem graphs, and struggles to find any such structural patterns in the random graphs of C and D.}
						\label{fig:pcd_example_dbm_vs_random}
					\end{center}
				\end{figure}

				Although the PCD algorithm can calculate global solutions for QUBOs of arbitrary size and structure, it was developed particularly for application to problem graphs with time structure. A common example is the Job Shop Scheduling problem, wherein penalties captured in the problem QUBO arise from temporal constraints, e.g. two jobs cannot be scheduled at the same time on the same machine. The PCD algorithm makes a series of cuts along the temporal axis of the problem graph, where each cut removes edges that intersect the cut. This process is repeated until each of the resulting disconnected graph components, representing subproblems of the original QUBO, is directly embeddable on an available quantum annealing device. This approach maximally preserves the initial temporal structure of the QUBO into each subproblem by attempting to find an appropriate hardware embedding after each additional graph cut. To determine where to cut the problem graph without needing any explicit structural information a priori, the PCD algorithm takes two primary steps:

				\begin{enumerate}
					\item \textit{Spring Layout algorithm.} To organize the graph, the spring layout algorithm of NetworkX \cite{networkx}, based off the force-directed Fruchterman-Reingold Algorithm \cite{Fruchterman1991}, is used to generate an initial 2D-coordinate pair for each graph node. The algorithm determines the 2D-coordinates through a simulation process, wherein the graph nodes undergo Brownian motion in an effort to diffuse as much as possible. Nodes connected by an edge are bound by a constraining force causing the overall graph layout to converge, after many simulation steps, such that nodes sharing many mutual connections gravitate towards each other. As stated in \cite{Fruchterman1991}, the exact convergence criteria for the number of outer loop iterations is a complex question, but heuristically we have seen that 1000 simulation iterations provides efficient results.
					\item \textit{Principal Component Analysis (PCA).} After the initial graph layout has been calculated using the spring layout algorithm, we apply PCA to the 2D-coordinates of all the nodes in the graph, to ensure that we are slicing the graph along the primary axis. We then use these components as the final graph coordinates; see Figure \ref{fig:pcd_example_dbm_vs_random} for examples of final graph coordinates generated from 4 different input graphs. After this transformation, the problem graph is partitioned, moving from left to right along the graph coordinates to determine subgraphs. If a subgraph cannot be embedded due to size or connectivity issues, it is continually sliced along the primary axis until an embedding of each of its components is found. These sub-QUBOs, defined by these embeddable subgraphs, are then sent to the quantum annealer.
				\end{enumerate}
				
				After generating solutions for all subproblems using the quantum annealing device, the PCD algorithm applys a greedy recomposition algorithm to combine all subproblem solutions into a global solution.

			\subsubsection{Freeze-and-Anneal (FA):} 
				FA is a novel heterogeneous optimization technique that extends the authors' prior work \cite{Bass2018}. The previous iteration of this algorithm was designed specifically to solve the SCA problem described in Section \ref{sec:Satellite}, while the FA algorithm presented in this work has been generalized and is capable of generating solutions for arbitrary QUBO problems.

				FA begins by using a genetic algorithm (GA) \cite{Goldberg_1988, Holland_1975} to evolve a population of potential solutions to the QUBO problem. The size of this population and the number of generations for which the GA will evolve are parameters explored in this work. Specifically, we experiment with the former set to 250 and 500, and the latter set to 10, 50, and 100. The ``fitness" of each solution is dependent on the corresponding energy of the QUBO for that solution; lower energy implies higher fitness. Because a solution is simply a collection of values of the QUBO's binary variables, each can easily be represented as a bit string (e.g. 00101001). After the population is allowed to evolve for a pre-specified number of generations, we iteratively perform the ``freeze" step to identify consistent variables. In this step, we analyze the population of all of the final bitstrings, looking for which variable is the most well-predicted. This per-variable frequency of a value (0 or 1) is weighted by the population-wide frequency of the respective value. 
			
				Each time a variable is frozen, the QUBO problem is modified by removing that variable. Any variables that couple with the frozen variable have their qubit biases increased by the respective coupling strength if the variable is frozen to ``1", or left unchanged if the variable is frozen to ``0". The freezing process steadily reduces the size of the QUBO, until the size of the remaining QUBO is small enough to be directly embeddable on the available quantum annealing device. This reduced QUBO consists of the binary variables that are the ``most difficult" for the classical technique to evaluate, and is assumed to be the ``hardest part" of the larger problem. The rationale behind this algorithm is that the quantum advantage can best be achieved by focusing on the hardest part of the problem. The solution returned by the quantum annealer is then combined with the frozen values to return the final combined solution.

			\subsubsection{Qbsolv:} 

				The open-source Qbsolv algorithm, outlined in \cite{Booth2017}, was initially created for the purpose of solving QUBO problems too large to natively fit on quantum annealing hardware. The Qbsolv algorithm leverages a quantum annealer's ability to heuristically search large solution spaces, while also accounting for the annealer's lack of precision through the use of a more precise, local minimization algorithm (namely, Tabu Search). In this work, analysis of Qbsolv was done so using Version 0.2.10 of the algorithm. We explore a single Qbsolv parameter, $NumRepeats$  - the number of main loop iterations that need to pass with no solution improvement before the algorithm is terminated, which we assign values of 1, 3, 10, 30, and 50.

			\subsubsection{Iterative Centrality Halo (ICH):} 
				The Iterative Centrality Halo (ICH) algorithm implemented in this work is an adaptation of the previous Core-Halo (CH) work of \cite{Chapuis2019}, in which CH-partitioning (originally introduced in \cite{Djidjev2016}) is applied to decomposing the problem graph of a QUBO representation of the Maximum Clique problem. The primary algorithmic steps of the ICH algorithm are as follows:
				
				\begin{enumerate}
					\item Define $max\_nodes$ as the size of the largest fully-connected problem graph that can be embedded onto your solver (on the 2000Q, this is 65).
					\item Given a QUBO $Q$ with problem graph $G$, let $S$ be a global solution vector with values of all variables initialized to 0. 
					\item Let $C$ be the node of highest degree in $G$, which we define as the ``central node".
					\item Let $H$ be the set of nodes connected to $C$. If $|H| > max\_nodes - 1$, randomly remove nodes from $H$ until $|H| = max\_nodes - 1$.
					\item Let $Q_{sub}$ be the sub-QUBO defined on $C$ and the nodes within $H$. Find an embedding of $Q_{sub}$ onto the quantum annealing hardware and solve.
					\item Set the values of $Q_{sub}$'s corresponding variables in $S$ to the values returned by the device.
					\item Propagate, through $Q$, the influence of setting these variables by modifying the qubit bias of each variable connected to a variable set to 1 by the coupling strength between these variables.
					\item Remove $C$ and the nodes within $H$ from $G$.
					\item Repeat steps (\rmnum{3}) - (\rmnum{8}) until the entire problem graph has been consumed.
				\end{enumerate}
				
				ICH departs significantly from the original algorithm in several ways, predominately in (1) the use of centrality to select central nodes which prioritizes nodes that likely have a more significant impact on the global solution, and (2) the iterative subgraph solution clamping process, which transfers information about the current global solution into the remaining QUBO.
				
			\subsubsection{Random Solver} 
				The intent of this work is not to form direct comparisons of QHOTs against state-of-the-art classical solvers, but rather to understand the benefits and detriments of using various QHOTs to solve particular problems. However, it remains necessary to establish a baseline comparison of classical performance. In terms of solution quality and constraint violation, the results returned from the random solver act as this baseline. When given a QUBO problem to solve, the random solver simply returns a random solution, i.e. all problem variables are assigned a random binary value (0 or 1). 

	\section{Results}\label{sec:results}
	
		\subsection{Experimental Set-up} \label{experimental_setup}
			The goal of this effort was to exhaustively compare the performance of various QHOT solvers on various types of optimization problems. In total, 104 unique problem instances were generated and solved (88 DBG, 14 TSP, 1 SCA, and 1 MWP). Problem graph characteristics, such as graph size (number of logical qubits) and edge density (ratio of the number of edges that do exist to the number of edges that could exist), are recorded for each problem. Additionally, 12 unique QHOT instances were used to solve each problem (1 PCD, 4 FA, 5 Qbsolv, 1 ICH, and 1 Random). Note that mutliple versions of the FA and Qbsolv algorithms are tested because we chose to explore the parameter space of these algorithms. Through the remainder of this section, we refer to an ``experiment" as a combination of (1) the type of problem being solved; (2) the parameter values that were used to generate the problem instance; (3) the QHOT being used to solve the problem; and (4) the parameter values of the QHOT. A complete list of all experiments conducted in this effort, as well as various performance metrics captured during the execution of each, are provided in the supplemental data.
		
			All experiments used the D-Wave 2000Q (maintained by NASA/USRA) as a quantum sub-processor. Experiments in which the DBG, TSP, and MWP problems were solved leveraged an Amazon Web Services (AWS) instance (c5d.xlarge) for classical processing. Due to data restrictions, experiments in which the SCA problem was solved leveraged the following classical processor: Intel Core i5-5300 @ 2.3 GHz. As such, time-based metrics for the SCA experiments are not comparable to that of other problems explored in this work.

		\subsection{Evaluation Criteria} \label{evaluation_criteria}
		To assess QHOT performance, the following metrics were primarily used:
		
		\begin{enumerate}			
			\item \textit{Total run-time}. Time-to-solution can be important for practical applications, and is essential for those that require near ``real-time" results. While any method leveraging cloud-based quantum computing will execute at far from ``real-time" speeds, we chose a ``maximum time threshold" to simulate some arbitrary time constraint that an end user may have for their application. If a QHOT experiment eclipsed this threshold, which naturally happens as problem sizes scale up, the experiment would be halted. In the interest of time, our experimental threshold was set at 30 minutes, which allowed the QHOT solvers to generate solutions for most experiments while avoiding extremely long run-times. 
			
			\item \textit{Relative run-time}. In addition to observing which experiments were able to complete within this threshold, it is also interesting to record the proportional time spent on \textit{classical}, \textit{embedding}, or \textit{quantum} tasks. These refer (respectively) to:
			\begin{enumerate}
				\item Computational time spent on the classical outer-loop parts of a given QHOT;
				\item The time spent using the heuristic problem embedding finding tool, minorminer \cite{embeddingz}, to map sub-QUBO problems onto the quantum annealing device; and 
				\item The time spent from submission of a sub-QUBO (and embedding) to the device until a solution response is returned (including time associated with network latency, the device's queue, and the annealing process).
			\end{enumerate}
			As the different QHOTs employ unique classical algorithms, handle embeddings in various ways, and query the quantum solver to a different extent, a breakdown of computational time across the three categories is useful for detecting computational bottlenecks and general trends for different QUBO problem graphs.
			
			\item \textit{Solution quality}. All QUBO problem terms are generated from problem constraints and an objective function. For each problem described in Section \ref{concrete_qubo_problems_and_generators}, \textit{solution energy} refers to the energy of a given solution, which is optimal when minimized. The number of \textit{broken constraints} refers to how many problem-specific constraints a given solution violates. For example, a TSP solution that does not visit all vertices exactly once violates one or more constraints that define a feasible solution. 
			
			For each problem (except the MWP), a method for recording and reconciling broken constraints, thereby ``fixing" an infeasible solution (usually, by some form of greedy post-processing), was implemented. We define the quality of a ``fixed" solution, referred to as \textit{fixed solution quality}, uniquely for each problem type, since a fixed solution is feasible and can therefore be assigned a problem-relevant measure of quality. For the DBG, TSP, SCA, and MWP problems, the fixed solution quality is: a generalized reward value, the path length, the total sub-constellation coverage, and a measure of the value of completed jobs, respectively. Note that TSP fixed solution quality is better when minimized (i.e. shortest route is ideal), while fixed solution quality for DBG, SCA, and MWP is better when maximized.
			
		\end{enumerate}

		\subsection{QHOT Parameter Testing} \label{parameter_testing}
			Since the ultimate goal of this research was to evaluate and compare the performance of various QHOTs, it was imperative that we understood the performance of the QHOTs within their own parameter spaces. By conducting this analysis, it allowed us to select the ``best" combination of parameter values for a given QHOT to use as a benchmark when comparing against the other techniques. In this work, the two solvers with variable parameters (FA and Qbsolv) were tested.
			
			\begin{figure}[H]
				\centering
				\includegraphics[width=0.7\linewidth]{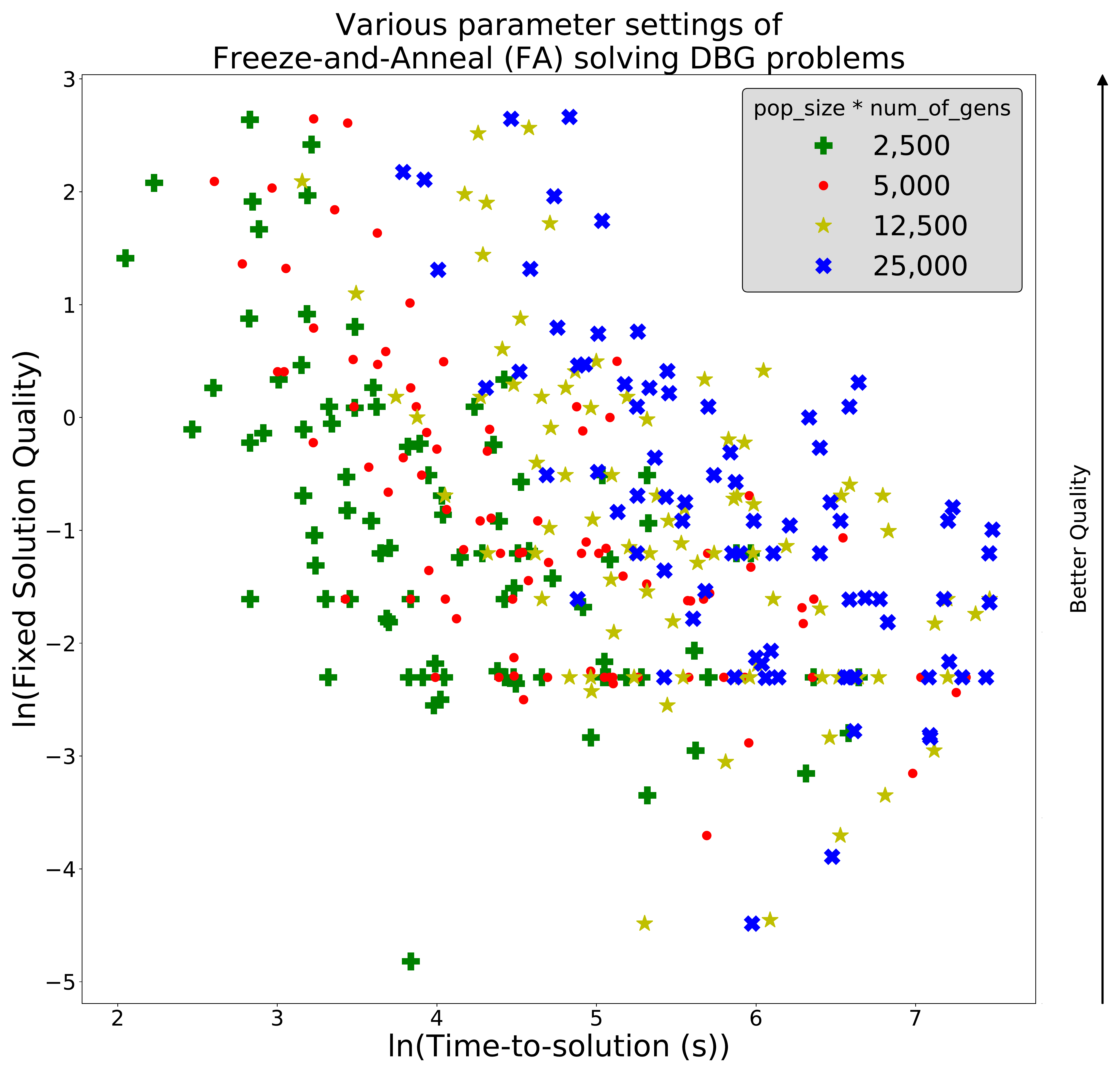}
				\caption{Comparing time-to-solution and fixed solution quality for various parameter settings of the FA algorithm solving DBG problems.}
				\label{fig:FA_solution_quality_and_timings}
			\end{figure}
			
			Figure \ref{fig:FA_solution_quality_and_timings} highlights a clear relationship between the total run-time of the FA algorithm and the product of the two parameters: number\_of\_generations and population\_size. This increase in run-time is expected and results directly from an increase in classical computational time; increasing the number\_of\_generations increases the overall number of classical iterations of the GA, while increasing population\_size similarly requires the evaluation of a larger number of solutions per iteration. Analyzing the fixed solution quality results, there is no clear strong performance advantage in choosing parameters that increase FA run-times in our experiments on DBG problems. Therefore, for the FA experiments, we chose to use the parameter values that resulted in the fastest run-times - that is, population\_size set to 250 and number\_of\_generations set to 10.
			
			\begin{figure}[H]
				\centering
				\includegraphics[width=0.7\linewidth]{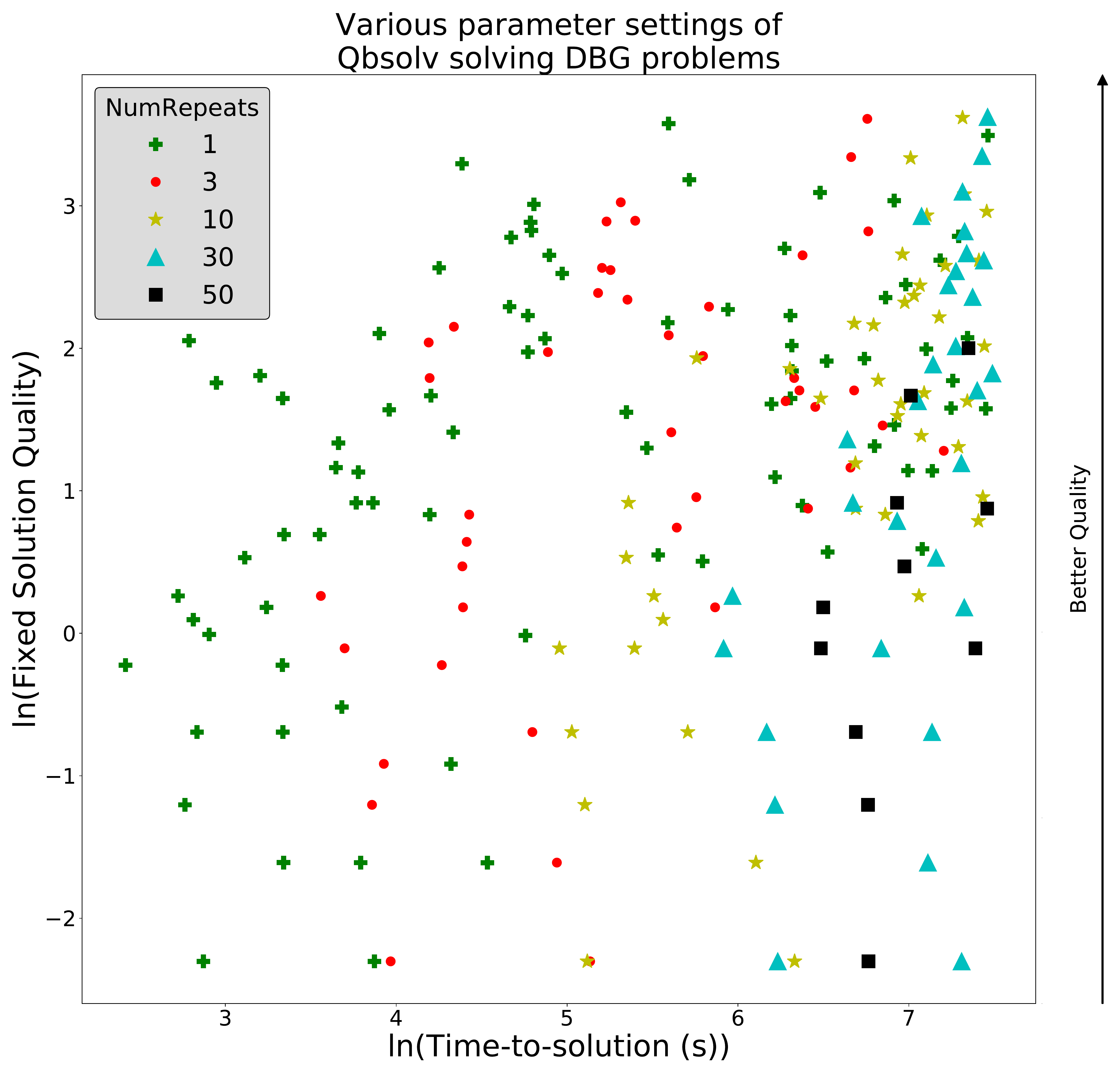}
				\caption{Comparing time-to-solution and fixed solution quality for various parameter settings of the Qbsolv algorithm solving DBG problems.}
				\label{fig:Qbsolv_solution_quality_and_timings}
			\end{figure}
			
			A similar trend was observed in Figure \ref{fig:Qbsolv_solution_quality_and_timings} in terms of the relationship between run-time and fixed solution quality. As NumRepeats increases, time-to-solution increases generally, with marginal improvements in quality being observed only when NumRepeats was set to 50 (the current default value for this parameter). Doing so, however, greatly reduced the number of DBG and TSP problems solvable and made the MWP and SCA problems unsolvable within the given time threshold. Therefore, we set NumRepeats to 1 for all Qbsolv experimental comparisons.

		\subsection{Performance of QHOTs}
		
\begin{figure}[H]
				\centering
				\includegraphics[width=1\linewidth]{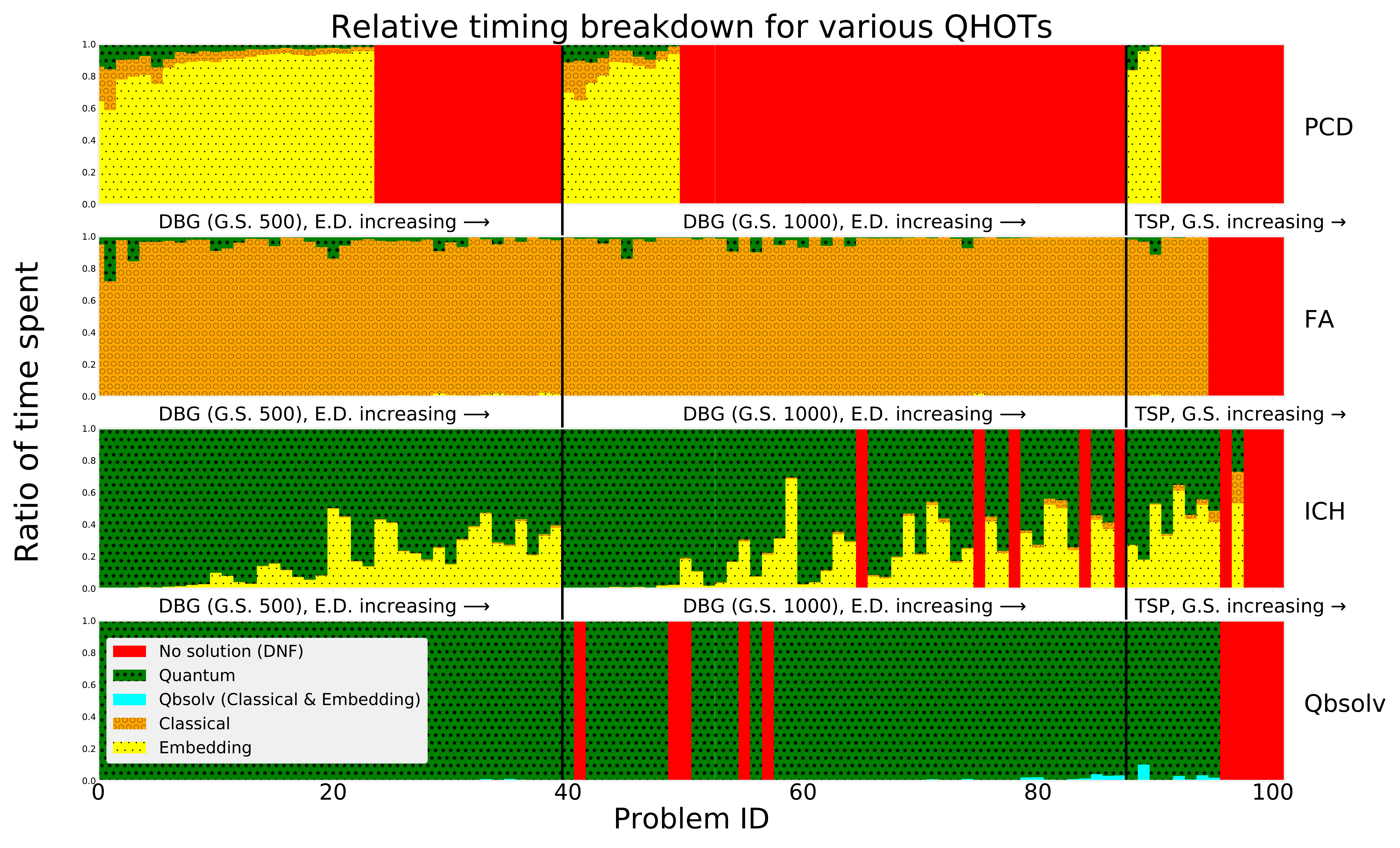}
				\caption{Relative timing breakdown for various QHOTs on DBG and TSP problems. Bars represents unique problems being solved and are arranged along the x-axis in the following order: DBG problems with Graph Size (G.S.) = 500 and Edge Density (E.D.) increasing, DBG problems with G.S. = 1000 and E.D. increasing, and TSP problems with G.S. increasing. A black bar is used to visually separate these sets of problems.}
				\label{fig:timing_breakdowns}
			\end{figure}

			Using the FA and Qbsolv parameter values determined in Section \ref{parameter_testing}, we evaluated the timing and performance across each of our QHOTs. Figure \ref{fig:timing_breakdowns} breaks down timing information, per solver, for all DBG and TSP problems solvable within the 30-minute upper-bound time threshold. 
			
			Interesting patterns emerge in comparing the timing breakdowns of the PCD, FA, Qbsolv, and ICH solvers, which were predominately dominated by embedding time, classical time, quantum time, and a combination of both quantum and embedding time, respectively. These results are best explained as a direct result of algorithm design: 
			\begin{enumerate}
				\item The majority of FA's run-time is spent classically optimizing via the GA before finally submitting a single reduced QUBO to the quantum annealer.
				\item PCD was designed to attempt to find an embedding of a sub-problem, further slicing this sub-problem along the primary axis if an embedding was not found. The algorithm used to find graph embeddings does not scale well with graph size and lacks adequate early dropout mechanisms - hence, the majority of PCD's time is spent searching for, and often failing to find, embeddings.
				\item Due to Qbsolv's black-box implementation (being an externally developed algorithm), it was difficult to classify how the algorithm's time was being computationally spent. Wrapping function calls made by the Qbsolv algorithm allowed for accurately recording quantum time. Unfortunately, we could not divide the remainder of Qbsolv's time into classical and embedding, so we report this non-quantum time as a combination of the two. The majority of Qbsolv's relative time is consumed by calls to the quantum device, with non-quantum time becoming noticeable for problems with larger size and/or higher edge density. It is likely that Qbsolv's non-quantum time is almost entirely classical, since the algorithm is designed to reuse a single problem embedding.
				\item Due to ICH's relatively straightforward classical component, the majority of its time is spent embedding and solving sub-problems. Despite not reusing a single problem embedding (as Qbsolv does), the relative embedding time of ICH is considerably less than that of PCD. This is because ICH is designed such that an embedding is guaranteed to exist for any sub-problem to be solved; therefore, no time is wasted waiting for the heuristic embedding finder to fail. 
			\end{enumerate}
		
			As expected for larger problem graphs with greater size and edge density (e.g. larger TSP problems), all QHOTs eventually struggled to find solutions within the time threshold. In terms of overall experiment completion, the FA, Qbsolv, and ICH algorithms were able to solve almost all of the same problems within the time threshold. 
			
			\begin{figure}[H]
				\centering
				\includegraphics[width=0.95\linewidth]{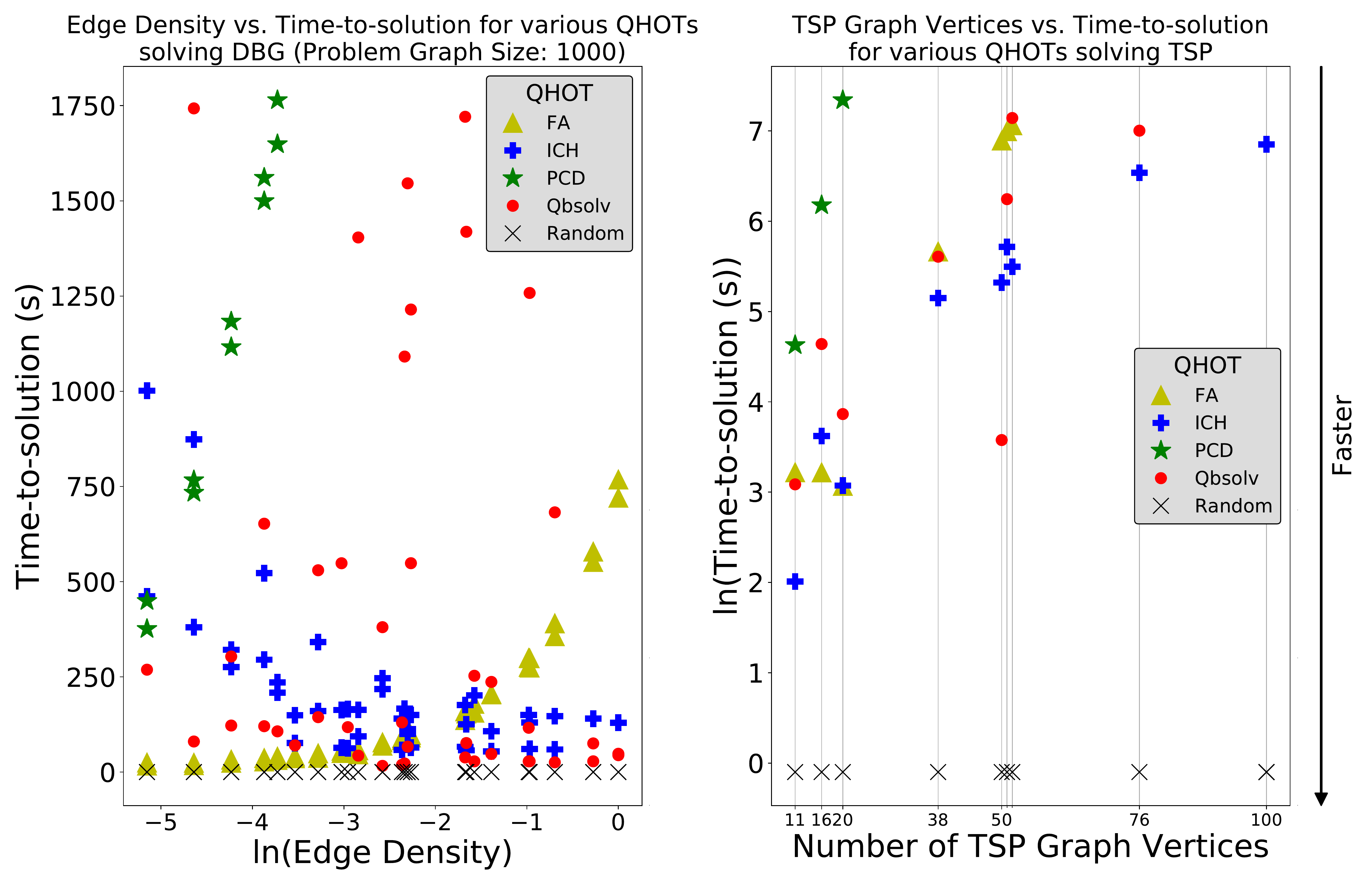}
				\caption{Comparing time-to-solution against both edge density and problem graph size for various QHOTs solving DBG (1000-node) and TSP problems. Similar results were observed for DBG problems of size 500.}
				\label{fig:TTS}
			\end{figure}
			
			In terms of time-to-solution for DBG and TSP, the results shown in Figure \ref{fig:TTS} echo that of Figure \ref{fig:timing_breakdowns}. As expected, all QHOTs require more time as problem size increases. In particular, PCD scales poorly as problem size and edge density increase. The same is true, to a lesser degree, for FA, which requires more time as edge density increases because graphs with higher edge density require more classical processing to propagate information of frozen qubits to adjacent variables. Conversely, ICH requires less time as edge density increases. Given a central node (as defined in the ICH algorithm), a higher overall problem graph edge density implies that that central node has closer to $max\_nodes$ neighbors. Therefore, while sub-problems are larger, on average, the total number of sub-problems to be solved is fewer, thus resulting in a decrease in overall solving time. Lastly, Qbsolv's time-to-solution, as opposed to that of the other QHOTs, is the least correlated with edge density and graph size. This is likely due to the fact that randomness is fundamental to the Qbsolv algorithm. While the same can be said about FA with its incorporation of a GA, the quality of randomly generated solutions has a direct effect on Qbsolv's termination and, therefore, its run-time. 
			
			\begin{figure}[H]
				\centering
				\includegraphics[width=0.95\linewidth]{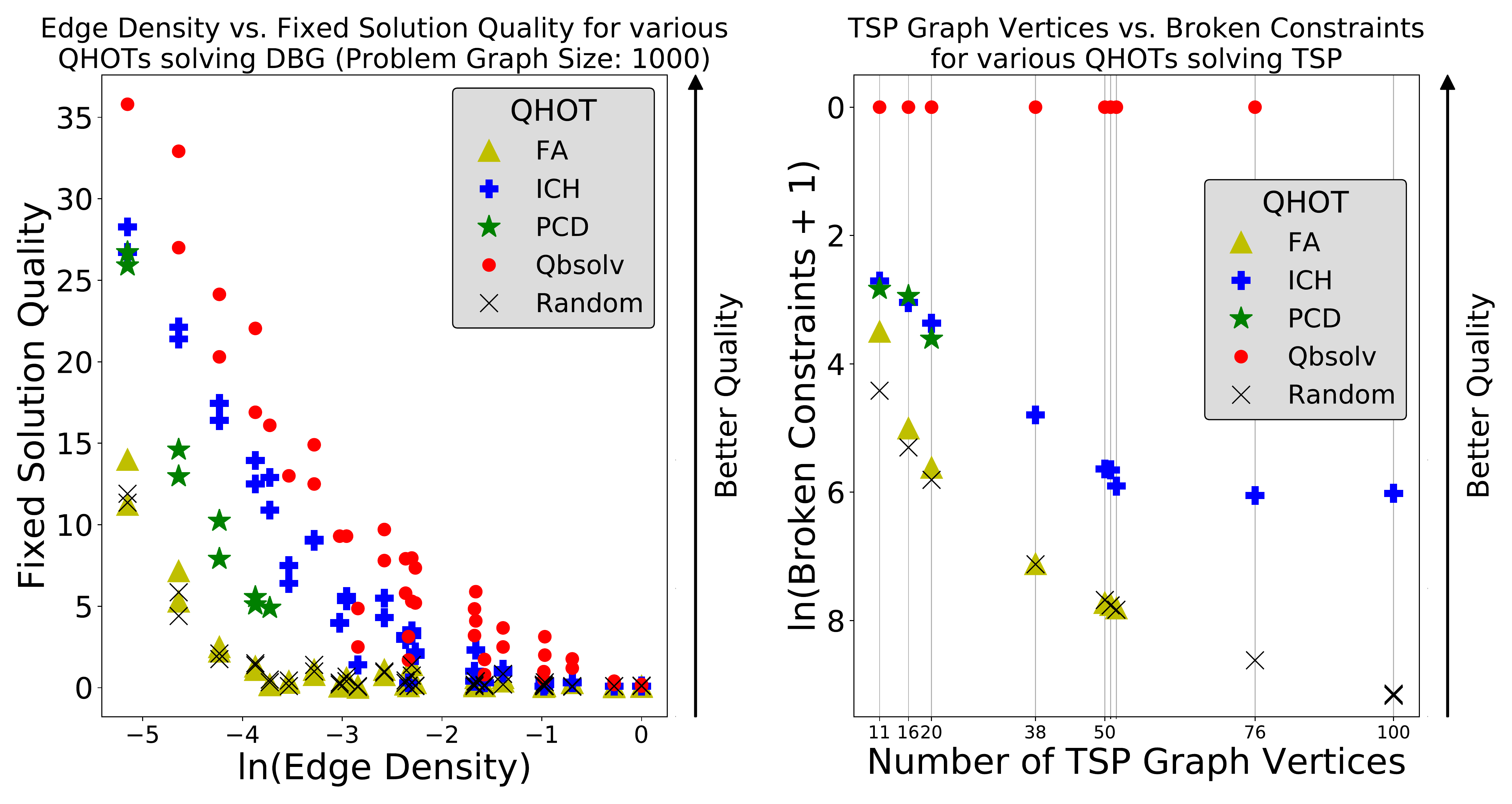}
				\caption{Comparing solution quality against both edge density and problem graph size for various QHOTs solving DBG (1000-node) and TSP problems. Similar results were observed for DBG problems of size 500.}
				\label{fig:qual}
			\end{figure}
			
			In terms of the quality of solutions to the DBG and TSP problems shown in Figure \ref{fig:qual}, Qbsolv produced the best solutions for every problem that it was able to solve within the time limit. This is likely because the number of broken constraints a solution has is directly correlated with solution quality, and the solutions that Qbsolv found, while not necessarily optimal, never had broken constraints. This ability to always return viable solutions to TSP problems can be seen in the righthand plot of Figure \ref{fig:qual}, though it is also the case for all DBG problems tested. Solutions found using the ICH algorithm typically had the second best quality, behind Qbsolv. PCD outperformed both the FA and Random solvers, but only produced solutions for problems with limited edge density and size. Lastly, of the non-random QHOTs evaluated in this work, FA produced the lowest quality solutions to the DBG and TSP problems, with quality comparable to, though slightly higher than, that of random solutions.
			
			Also evident from our results is the difficulty, across all solvers, in finding solutions to TSPs with 10,000+ qubits (i.e. 100+ vertices to visit) in a timely fashion. Of the 6 100-vertex TSPs (1 Random, 5 from TSPLIB), 1 was solved in 30 minutes, and by only a single non-random QHOT.
			
			\begin{table}[H]
				\begin{center}
					\resizebox{\linewidth}{!}{
						\begin{tabular}{|c|c|c|c|c|c|}
							\hline 
							\textbf{Problem (Size, Edge Density)} & \textbf{Solver} & \textbf{Total time (s)} & \textbf{Solution Energy} & \textbf{Broken Constraints} \\
							\hline
							\multirow{5}{*}{SCA (1013, 1)} & ICH & \num{260.295} & \num{56794709} & \num{1674} \\
							& FA & \num{249.714} & \num{458850157} & \num{75413} \\
							& PCD & \num{1105.106} & \num{962418755} & \num{331662} \\
							& Qbsolv & \num{139.148}\color{red}* & \num{1949602}\color{red}*& \num{0}\color{red}* \\
							& Random & \num{0.079}	& \num{470288031} & \num{79067} \\
							\hline
							\multirow{5}{*}{MWP (2280, 0.0026)}& ICH & \num{10538} & $-1.356 \times 10^5$ & N/A \\
							& FA & \num{1350.49} & \num{-67160.10} & N/A \\
							& PCD & \num{849.87}  & $-1.313 \times 10^5$ & N/A \\
							& Qbsolv & \num{40.03}\color{red}* & $-1.357 \times 10^5$\color{red}* & N/A \\
							& Random & \num{0.0029} & \num{-53025.93} & N/A \\
							\hline
							\multicolumn{5}{|c|}{\color{red}{*} indicates best performance}\\
							\hline
					\end{tabular}}
					\caption{Experimental results for all QHOTs on MWP and SCA problems.}
					\label{tab:sat_mwp_results} 	
				\end{center}
			\end{table}
		
			QHOT performance results when solving the MWP and SCA problems are shown in Table \ref{tab:sat_mwp_results}. Similar to both DBG and TSP problems, Qbsolv outperforms all other QHOTs in terms of both time-to-solution and solution energy, for the MWP and SCA problems. ICH places second for each of the aforementioned categories, except MWP time-to-solution in which it places last. This is likely due to the MWP having a very sparse problem graph, which ICH struggles with.

	\section{Conclusion and Future Work}\label{sec:conclusion}
	
		The QHOTs presented in this work displayed the capability of solving problems consisting of hundreds and even thousands of problem variables, which are considerably larger than the problem graphs directly embeddable onto current generation quantum devices. Even larger problems may be solved and solution quality increased by analyzing some key takeaways from this work and extending current and future QHOTs to maximize their utility. 
		
		\textit{Sub-QUBO information propagation}. For the solvers that produced the highest quality solutions, namely ICH and Qbsolv, sub-QUBO solutions are iteratively used to update the remaining global QUBO problem while determining the global solution. This ``information propagation" is not harnessed, in any capacity, by the PCD and Random solvers, which produced solutions of worse quality than ICH and Qbsolv on all problems evaluated in this work. A modified version of this information propagation is implemented by the FA algorithm, which produced solutions with qualities similar to the Random solver. However, while Qbsolv and ICH propagated information gained from sending sub-QUBOs to the quantum annealer, FA propagated only classically-determined information gained after the GA had optimized over the entire solution space.
		
		\textit{Reuse of static embeddings}. As evidenced by the PCD and ICH results of Figure \ref{fig:timing_breakdowns}, constantly recalculating hardware embeddings becomes a major computational bottleneck as problem size and edge density increase. Significant temporal resources are consumed when heuristically searching for an embedding. This is especially true for sub-problems that are impossible to embed due to their size or edge density because the algorithm that minorminer implements scales poorly as problem graph size increases and often consumes a significant amount of time before returning an indication of failure in the event that an embedding cannot be found. To avoid this issue, one pragmatic approach for use in future QHOTs is to pre-calculate the hardware embedding for the largest fully-connected problem graph that can be embedded onto an available quantum device (this is 65 for the 2000Q). Rather than repeatedly spending time searching for embeddings, this single embedding can be reused by trivially mapping any sub-problem with equal or lesser size to it. Qbsolv takes an approach similar to this, which explains the minimal non-quantum time usage. One drawback of this approach is that sub-problems of a greater size with less-than-fully-connected problem graphs may be embeddable, but are further reduced. However, alleviating this drawback introduces the risk of wasting time searching for an embedding that does not exist which, given current embedding finder heuristics, greatly reduces the temporal efficiency of QHOTs that accept this risk (e.g. PCD). Additionally, general improvements in the heuristic embedding search algorithm (increased speed, more advanced drop-out for problems that cannot be embedded, etc.) will naturally result in improvements in time-to-solution for QHOTs that require recalculation of hardware embeddings. 

		\textit{Additional improvements} can be had by introducing these techniques (sub-QUBO information propagation and embedding reuse) to those QHOTs that do not leverage both. Additionally, improvements may arise from employing different classical optimization algorithms (i.e. simulated annealing, Monte Carlo methods, etc.) other than the GA (used by FA) and the Tabu search (used by Qbsolv). It is also likely that dynamically repairing broken constraints using either problem-specific or problem-agnostic information during algorithm execution, rather than after, will produce higher quality solutions, and is therefore a research area that should be explored. Beyond improving the QHOTs that we explored, there are certainly other decomposition techniques that should be evaluated. For example, the techniques used in \cite{Ajagekar2019} can be ported into our framework and generalized to solve arbitrary QUBO problems.
		
		Lastly, it has been shown that, when solution quality is solely considered, Qbsolv (with NumRepeats set to 1) is superior for solving both theoretical (DBG, TSP) and real-world (MWP, SCA) problems. However, Qbsolv's time-to-solution relies heavily on randomness, and therefore, Qbsolv is the most temporally unpredictable of the QHOTs explored in this work. Currently, the default value for Qbsolv's NumRepeats parameter is 50. Given the marginal improvements in solution quality that increasing NumRepeats has shown, it is likely that one can comfortably, in the interest of time, afford to decrease NumRepeats to a value dependent on the criticality of time-to-solution with respect to solution quality.

	\ack 
	This work was funded through the Air Force Research Laboratory, Contract No. FA8750-18-C-0167. The authors would like to acknowledge the support of the Universities Space Research Association, Quantum AI Lab Research Opportunity Program, Cycle 2, for providing us with remote access to a D-Wave 2000Q device. Additionally, the thoughtful reviews and suggestions from Duncan Fletcher are acknowledged.
	
	\newcommand{\newblock}{}
	\bibliographystyle{plain}
	\bibliography{hots}
	
\end{document}